\documentclass[a4paper,12pt,abstract=true]{scrartcl}
\usepackage{authblk}
\usepackage{amsmath,amssymb}
\usepackage{bm}
\usepackage{graphicx,color}
\usepackage[hidelinks]{hyperref}
\usepackage{url}
\usepackage{cite}

\usepackage[all,warning]{onlyamsmath}
\RequirePackage[l2tabu, orthodox]{nag}

\newbox{\ORCIDicon}
\sbox{\ORCIDicon}{\large
                  \includegraphics[width=0.8em]{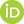}}

\begin{document}
\titlehead{\hfill OU-HET-1182}
\title{Rare Leptonic Processes Induced by Massless Dark Photon}

\author[1]{Xiaolong Deng}

\author[2]{Florentin Jaffredo\,\href{https://orcid.org/0000-0002-7673-9448}
                               {\usebox{\ORCIDicon}}
           \thanks{Email: \texttt{florentin.jaffredo@pi.infn.it}}}

\affil[1]{Department of Physics, Graduate School of Science, 
          Osaka University, Toyonaka, Osaka 560-0043, Japan} 

\author[1]{Minoru~Tanaka\,\href{https://orcid.org/0000-0001-8190-2863}
                               {\usebox{\ORCIDicon}}
           \thanks{Email: \texttt{tanaka@phys.sci.osaka-u.ac.jp}}}

\affil[2]{INFN, Sezione di Pisa, Largo Bruno Pontecorvo 3, I-56127 Pisa, Italy}

\date{\normalsize June 16, 2023}

\maketitle

\begin{abstract}
We introduce a dark photon considering a U(1) gauge extension of the standard model in particle physics.
Provided that the extra U(1) symmetry is unbroken, the dark photon is massless and has no coupling to the standard electromagnetic current. 
Higher-dimensional operators describe interactions of the massless dark photon with particles in the standard model. 
We investigate the interactions of the massless dark photon with charged leptons via dipole operators, mainly focusing on the lepton family-violating processes.
We present an improved constraint in the polarized two-body muon decay and a set of new bounds in tau decays.
We also examine possible lepton family-violating signals of the massless dark photon in future lepton colliders.
\end{abstract}

\section{Introduction\label{Sec:Introduction}}
One of the minimal extensions of the standard model (SM) is the U(1) extension in the gauge sector. 
Provided that the SM particles transform nontrivially under the extra U(1) gauge symmetry, there are multiple choices for anomaly-free U(1) charges, such as $B-L$, $L_\mu-L_\tau$, see, \textit{e.g.}, Refs.~\cite{Correia:2019woz,Costa:2020krs,Davighi:2022dyq}. In the present work, as an alternative possibility, we consider the case that no SM particles are charged under the extra U(1) gauge symmetry~\cite{Holdom1986a}. The gauge boson associated with such a U(1) symmetry is often called a dark photon. It is then natural to suppose the existence of a dark sector that consists of particles that are charged under the extra U(1) and not under the SM gauge group. 

The dark photon (denoted by $\gamma'$) and the dark sector may interact with the SM sector through the gauge-invariant kinetic mixing, $\propto F'_{\mu\nu}F^{\mu\nu}$, between the dark U(1) field strength ($F'_{\mu\nu}$) and $\text{U(1)}_\text{em}$ ($F_{\mu\nu}$).
Heavy particles charged under both the dark U(1) and $\text{U(1)}_\text{em}$ induce such a mixing by loop effects~\cite{Holdom1986a}.
An appropriate field redefinition leads to canonical kinetic terms for two U(1) gauge fields.
If the dark U(1) is broken and the dark photon is massive accordingly, a mass mixing arises
and the dark photon is coupled to the SM current through the mass mixing in this canonical basis.

In the case of a massless dark photon, two U(1) gauge bosons have the same quantum number.
The physical states are identified by their interactions.
Since we identify the ordinary photon as a massless gauge boson coupled to the SM electromagnetic current, we define the dark photon as a massless gauge boson that is not coupled to the SM electromagnetic current.
Consequently, the massless dark photon interacts with the SM sector via higher-dimensional operators.

The operators of the lowest dimension are dipole operators of dimension five~\cite{Dobrescu2005a}.
In this work, we concentrate on the charged-lepton dipole interactions described by the following lagrangian,
\begin{align}\label{Eq:LdipoleRL}
\mathcal{L}_\text{dipole}=-\frac{1}{2}\sum_{f',f=e,\mu,\tau}
 \bar f'(D_R^{(f'f)}P_R+D_L^{(f'f)}P_L)\sigma^{\alpha\beta}f F'_{\alpha\beta}\,,
\end{align}
where $D_{R,L}^{(f'f)}$'s are dipole coupling constants of dimension $-1$, and $P_{R/L}=(1\pm\gamma_5)/2$. 
The hermiticity of the lagrangian implies
\begin{align}
D^{(ff')}_R=D^{(f'f)*}_L\,.
\end{align}
We note that both the family-conserving and family-violating dipole interactions are
gauge-invariant and thus allowed. 

The above lagrangian is an effective lagrangian at low energies.
The information of UV theory is encoded in the dipole coupling constants.
We treat the dipole coupling constants as free parameters and investigate experimental constraints mainly focusing on lepton family-violating (LFV) processes.
Although we will briefly discuss some implications for a UV model,  apart from that, our analysis is model-independent.

The rest of the paper is organized as follows.
In Sec.~\ref{Sec:Decays}, we examine leptonic rare processes of muon and tau decays. 
We also present complementary arguments on lepton family-conserving processes in the same section.
In Sec.~\ref{Sec:Colliders}, we study possible dark photon signatures in proposed lepton collider experiments. 
Section~\ref{Sec:Summary} is devoted to discussions and summary.

\section{Leptonic rare decays\label{Sec:Decays}}
\subsection{Muon decays}
\subsubsection{\texorpdfstring{$\mu\to e\gamma'$}{mu to e gamma'}}
The decay rate of $\mu\to e\gamma'$ is given by
\begin{align}\label{Eq:Gmuegp}
\Gamma(\mu\to e\gamma')=
\frac{1}{16\pi}(|D_R^{(\mu e)}|^2+|D_L^{(\mu e)}|^2)m_\mu^3
                        \left(1-\frac{m_e^2}{m_\mu^2}\right)^3\,.
\end{align}
An experimental bound on the branching fraction of $\mu\to e+X$, where $X$ represents a missing massless boson, $\text{Br}(\mu\to e+X)<5.8\times 10^{-5}$~\cite{TWIST2015},  is employed in Ref.~\cite{FGL2020a}. In our notation we obtain 
\begin{align}
\frac{1}{\sqrt{|D_R^{(\mu e)}|^2+|D_L^{(\mu e)}|^2}}\ge 1.2\times 10^6\ 
\text{TeV}\,.\label{Eq:mutoegammaLimit}
\end{align}
In the following, we argue that the above bound is appropriate if $D_L^{(\mu e)}=0$, and improve it for the case of $D_L^{(\mu e)}\neq 0$.

The above experimental bound of muon decay is given by the TWIST Collaboration~\cite{TWIST2015}, which studied polarized $\mu^+$ decays.
We take the spin quantization axis being the positive $z$ direction and denote the polar angle of the $e^+$ momentum in the $\mu^+$ rest frame by $\theta$.
The decay distribution is given by
\begin{align}\label{Eq:dGmuegp}
\frac{1}{\Gamma(\mu^+\to e^+\gamma')}\frac{d\Gamma(\mu^+\to e^+\gamma')}
                                          {d\cos\theta}
=\frac{1}{2}\left(1-AP_{\mu^+}\cos\theta\right)\,.
\end{align}
where $\Gamma(\mu^+\to e^+\gamma')$ is the decay rate in Eq.\eqref{Eq:Gmuegp}, the asymmetry factor $A$ is defined by
\begin{align}
A:=-\frac{|D_R^{(\mu e)}|^2-|D_L^{(\mu e)}|^2}{|D_R^{(\mu e)}|^2+|D_L^{(\mu e)}|^2}\,,
\end{align}
and $P_{\mu^+}$ represents the $\mu^+$ polarization. 
In the present convention, $P_{\mu^+}\simeq -1$ in the TWIST experiment~\cite{TWIST2015}.
In Ref.~\cite{TWIST2015}, the TWIST Collaboration has analyzed three cases of angular distribution, $A=\pm 1, 0$.
Their result is summarized in Table \ref{Tab:TWIST} as well as the corresponding bounds on the dipole coupling constant.

\begin{table}[ht]
\caption{Upper bounds on the branching fraction of $\mu\to e+X$ for $A=\pm 1$~\cite{TWIST2015}, and corresponding bounds on the dipole coupling.}
\label{Tab:TWIST}
\centering
\begin{tabular}{c|ccc}
A  & -1 & 0 & +1 \\ \hline
Br ($\times 10^{-5}$) & 5.8 & 2.1 & 1.0\\
$(|D_R^{(\mu e)}|^2+|D_L^{(\mu e)}|^2)^{-1/2}$ [$10^6$ TeV] & 1.2 & 1.9 & 2.8
\end{tabular}
\end{table}

We note that the case of $A=-1$ ($D_L^{(\mu e)}=0$) exhibits the same angular distribution as the SM three-body muon decay at the endpoint.
As shown in Table \ref{Tab:TWIST}, the experimental bound for the $A=-1$ case is the weakest because the SM background is the most severe.
We have found better bounds for the cases of $A=0$ and $+1$.

\begin{figure}[ht]
 \centering
 \includegraphics[width=0.6\textwidth]{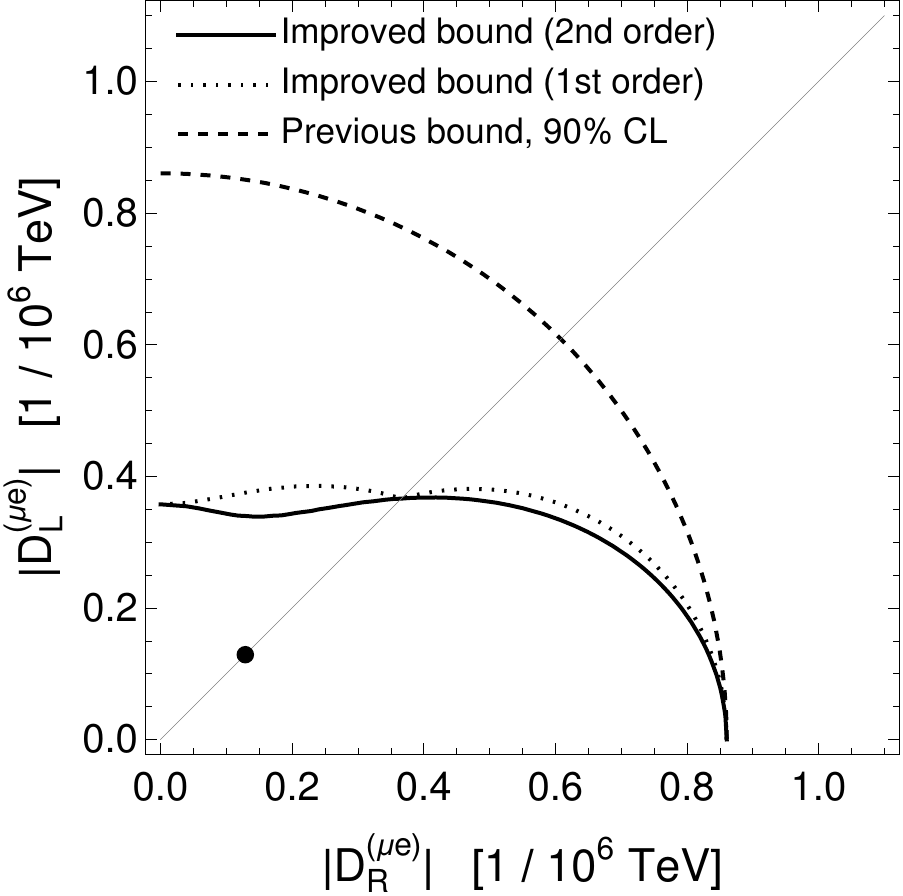}
 \caption{Improved bounds interpolating the TWIST result~\cite{TWIST2015}. Solid (Dotted): the second (first) order interpolation. Dashed: the previous bound ($A=-1$ only).
 The black dot represents the bound obtained from the familon search~\cite{Jodidio1986}.
 (See the main text for details.)}
 \label{Fig:TWISTbound}
\end{figure}

Interpolating the TWIST data for $A=\pm 1, 0$, we obtain improved bounds in the $|D_R^{(\mu e)}|$-$|D_L^{(\mu e)}|$ plane as depicted in Fig.~\ref{Fig:TWISTbound}. 
The solid (dotted) line represents the 90\% CL bound with the second (first) order interpolation, where the higher-order polynomial is used in order to estimate the interpolation error.
The dashed line is the previous bound ignoring the angular-distribution dependence, namely the bound in Eq.\eqref{Eq:mutoegammaLimit}.

An upper bound on the branching fraction of the muon decay into an electron and a familon~\cite{Wilczek1982a} (denoted by $f$) is given in Ref.~\cite{Jodidio1986}, $\text{Br}(\mu\to ef)<2.6\times 10^{-6}$.
This bound applies to the case of $A=0$ ($|D_R^{(\mu e)}|=|D_L^{(\mu e)}|$) in $\mu\to e\gamma'$ since the familon is isotropically emitted in the muon rest frame. 
The black dot on the diagonal line in Fig.~\ref{Fig:TWISTbound} represents this bound,
\begin{align}\label{Eq:Jodidio}
\frac{1}{\sqrt{|D_R^{(\mu e)}|^2+|D_L^{(\mu e)}|^2}}>5.5\times 10^6\ \text{TeV}\,.
\end{align}

\subsubsection{\texorpdfstring{$\mu^+\to e^+e^-e^+$}{mu to eee}}
\begin{figure}[t]
 \centering
 \includegraphics[width=0.8\textwidth]{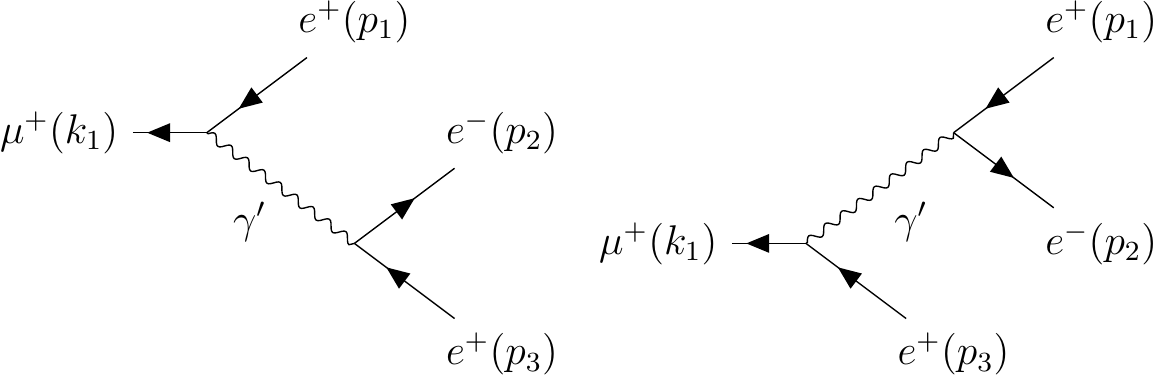}
 \caption{Diagrams involved in $\mu\to e^+e^-e^+$. Note that they are not equivalent.}
 \label{fig:diagrams:muto3e}
\end{figure}

In the presence of both family-conserving $(ee)$ and violating $(\mu e)$ couplings, the effective lagrangian $\mathcal{L}_\text{dipole}$ in Eq.~\eqref{Eq:LdipoleRL} causes the decay $\mu^+\to e^+e^-e^+$ .
Two distinct diagrams contribute to the process as depicted in Fig.~\ref{fig:diagrams:muto3e}.
We denote the amplitude of the left (right) diagram by $M_{1(2)}$.
Neglecting the electron mass, the squared amplitudes with the final spins summed, and the initial spin averaged
are given by
\begin{align}
\sum_\text{spin}|M_1|^2 &=
 (|D_R^{(\mu e)}|^2+|D_L^{(\mu e)}|^2)|D_R^{(ee)}|^2
 \{m_\mu^2(t+4u)-(t+2u)^2\}\,,\\
\sum_\text{spin}|M_2|^2 &=
 (|D_R^{(\mu e)}|^2+|D_L^{(\mu e)}|^2)|D_R^{(ee)}|^2
 \{m_\mu^2(u+4t)-(u+2t)^2\}\,,
\end{align}
where $t=(k_1-p_1)^2=(p_2+p_3)^2$, $u=(k_1-p_3)^2=(p_1+p_2)^2$, and we have used $D_R^{(ee)}=D_L^{(ee)*}$. 
The interference of the two amplitudes is expressed by
\begin{align}
&\sum_\text{spin}M_1 M_2^* =\sum_\text{spin}M_1^* M_2\nonumber\\
 &=\frac{1}{2}(|D_R^{(\mu e)}|^2+|D_L^{(\mu e)}|^2)|D_R^{(ee)}|^2
   \{2 m_\mu^2(t+u)-(t+2u)(u+2t)\}\,.
\end{align}
We note that the relative sign owing to the Fermi statistics is included.
Integrating over the phase space, we obtain the decay rate as
\begin{align}
 \Gamma(\mu^+\to e^+ e^- e^+)
 =\frac{1}{2 m_\mu}\int d\Phi_3\sum_\text{pol}|M_1+M_2|^2\,,\\
 =\frac{5m_\mu^5}{2048\pi^3}
   (|D_R^{(\mu e)}|^2+|D_L^{(\mu e)}|^2)|D_R^{(ee)}|^2\,,
\end{align}
where the three-body phase space is given by
\begin{align}
d\Phi_3=\frac{m_\mu^2}{128\pi^3}dz_1 dz_3\,,
\ z_i:=\frac{2k_1\cdot p_i}{m_\mu^2}\,.
\end{align}
The integration region is $0\leq z_1,z_3\leq 1$ and $z_1+z_3\geq 1$.

With the experimental constraint $\text{Br}(\mu^+\to e^+e^-e^+)<1.0 \times 10^{-12}$~\cite{SINDRUM1988}, we find
\begin{align}
\frac{1}{[(|D_R^{(\mu e)}|^2+|D_L^{(\mu e)}|^2)|D_R^{(ee)}|^2]^{1/4}} > 243\ \text{TeV}\,.
\end{align}

\subsection{Tau decays}
\subsubsection{\texorpdfstring{$\tau\to\ell\gamma'$}{tau to l gamma'}}
The search for $\tau\to \ell\gamma'$ ($\ell=\mu,e$)
constrains the corresponding parameters, $D_{R,L}^{(\tau\ell)}$, in Eq.~\eqref{Eq:LdipoleRL}.
The decay rate is given by Eq.~\eqref{Eq:Gmuegp} with an appropriate change of charged leptons.
With the recent experimental bounds given by the Belle II Collaboration, 
$\text{Br}(\tau\to \mu(e)+X)<0.59(0.94)\times 10^{-3}$ (95 \% CL)~\cite{BelleII2022a},
we obtain
\begin{align}
\frac{1}{\sqrt{|D^{(\tau\mu)}_R|^2+|D^{(\tau\mu)}_L|^2}}
 \ge 9.1\times 10^3\ \text{TeV}\,,\ 
\frac{1}{\sqrt{|D^{(\tau e)}_R|^2+|D^{(\tau e)}_L|^2}}
 \ge 7.2\times 10^3\ \text{TeV}\,.\label{Eq:tautoeormuLimit}
\end{align}

\subsubsection{\texorpdfstring{$\tau^+\to\ell^+\ell^-\ell^+$}{tau to l l l}}
Similar diagrams as $\mu^+\to e^+ e^- e^+$ describe $\tau^+\to\ell^+\ell^-\ell^+$ ($\ell=\mu,e$).
The decay rate is given by
\begin{align}
 \Gamma(\tau^+\to\ell^+\ell^-\ell^+)
 =\frac{5m_\tau^5}{2048\pi^3}
   (|D_R^{(\tau\ell)}|^2+|D_L^{(\tau\ell)}|^2)|D_R^{(\ell\ell)}|^2\,.
\end{align}
The experimental bound is $\text{Br}(\tau^+\to\ell^+\ell^-\ell^+)<2.1(2.7)\times 10^{-8}$ for $\ell=\mu(e)$~\cite{Hayasaka2010}.
The constraints on the relevant dipole couplings are
\begin{align}
\frac{1}{[(|D_R^{(\tau\mu)}|^2+|D_L^{(\tau\mu)}|^2)|D_R^{(\mu\mu)}|^2]^{1/4}}
> 13.1\ \text{TeV}\,,
\end{align}
and 
\begin{align}
\frac{1}{[(|D_R^{(\tau e)}|^2+|D_L^{(\tau e)}|^2)|D_R^{(ee)}|^2]^{1/4}}
> 12.3\ \text{TeV}\,,\label{Eq:tautoeeeLimit}
\end{align}

\subsubsection{\texorpdfstring{$\tau^+\to\ell^+\ell'^-\ell'^+$}{tau to l lp lp}}
Here, we examine other $\tau$ decay modes into three charged leptons ignoring the dipole couplings $D_{R,L}^{(\mu e)}$ since they are strongly constrained by the muon two-body decay discussed above.
The relevant process, $\tau^+\to\ell^+\ell^{'-}\ell^{'+}$ ($\ell\neq\ell'$), is described by one diagram. The decay rate is expressed as
\begin{align}
 \Gamma(\tau^+\to\ell^+\ell^{'-}\ell^{'+})
 =\frac{m_\tau^5}{1024\pi^3}
   (|D_R^{(\tau\ell)}|^2+|D_L^{(\tau\ell)}|^2)|D_R^{(\ell'\ell')}|^2\,.
\end{align}
The experimental bounds, $\text{Br}(\tau^+\to\mu^+e^-e^+)<1.8\times 10^{-8}$ and $\text{Br}(\tau^+\to e^+\mu^-\mu^+)<2.7\times 10^{-8}$~\cite{Hayasaka2010}, lead to the following constraints:
\begin{align}
&\frac{1}{[(|D_R^{(\tau\mu)}|^2+|D_L^{(\tau\mu)}|^2)|D_R^{(ee)}|^2]^{1/4}}
> 10.8\ \text{TeV}\,,\label{Eq:tautomueeLimit}\\
&\frac{1}{[(|D_R^{(\tau e)}|^2+|D_L^{(\tau e)}|^2)|D_R^{(\mu\mu)}|^2]^{1/4}}
> 9.77\ \text{TeV}\,.\label{Eq:tautomumueLimit}
\end{align}

\subsection{Other constraints on lepton family-conserving couplings}
We briefly mention other constraints on family-conserving couplings, $D_{R,L}^{(\ell\ell)}$, and present constraints in our notation.

The emission process of the massless dark photon in stellar plasma provides an extra cooling mechanism in the stellar evolution. 
The combined data of cooling in white dwarves and red giants \cite{Giannotti2016a} result in~\cite{FGL2020a}
\begin{align}\label{Eq:SC}
|D_{R,L}^{(ee)}|\gtrsim 7\times 10^6\ \text{TeV}\ 
\text{(Stellar cooling)}.
\end{align}

The effective number of relativistic degrees of freedom at the era of the big bang nucleosynthesis (BBN) is given by $N_\text{eff}=2.878\pm 0.278$ \cite{Fields2020a}. 
Accommodating two degrees of freedom of the massless dark photon is unlikely in the presence of the three light neutrinos.
In order to dilute the contribution of the massless dark photon to the energy density of the universe at the BBN era, the decoupling temperature of the massless dark photon is required to be higher than the temperature of the QCD phase transition ($\sim 150 \text{MeV}$).
This requirement implies~\cite{Dobrescu2005a,FGL2020a}
\begin{align}\label{Eq:BBN}
|D_{R,L}^{(\ell\ell)}|\gtrsim 1\times 10^4\ \text{TeV}\ 
\ell=e,\mu\ \text{(BBN)}.
\end{align}

In addition to the above astrophysical and cosmological constraints, several laboratory constraints are studied in the literature \cite{FGL2020a}.
Apart from them, muonium hyperfine splitting (HFS) has a potential sensitivity to family-conserving couplings.
We find that the contribution of the massless dark photon to the HFS of muonium in the $n$S state is given by
\begin{align}
\Delta E_\text{HFS}(n\text{S})
=\left(\frac{2}{3}D_M^{(ee)}D_M^{(\mu\mu)}-\frac{1}{3}D_E^{(ee)}D_E^{(\mu\mu)}\right)
 \frac{1}{\pi}\left(\frac{Z\alpha m_r}{n}\right)^3\,,
 \end{align}
where the magnetic ($D_M^{(\ell\ell)}$) and electric ($D_E^{(\ell\ell)}$) dipole couplings are related to $D_{R/L}^{(\ell\ell)}$ by $D_{R/L}^{(\ell\ell)}=D_M^{(\ell\ell)}\pm i D_E^{(\ell\ell)}$, $Z=1$ is the charge of the antimuon, and $m_r=m_e m_\mu/(m_e+m_\mu)$ represents the reduced mass.
The experimental uncertainties at present are 710 Hz~\cite{MuSEUM2021a} and 53 Hz~\cite{Liu1999a} at zero and high magnetic field respectively, while the uncertainty in the standard model prediction is
271 Hz~\cite{Mohr2016a} or 516 Hz~\cite{Eides2019a} depending on the authors~\cite{KarshenboimKorzinin2021a}.
Assuming $|\Delta E_{HF}(1\text{S})/h|<500\ \text{Hz}$,
we find
\begin{align}
\left|\frac{2}{3}D_M^{(ee)}D_M^{(\mu\mu)}-\frac{1}{3}D_E^{(ee)}D_E^{(\mu\mu)}
\right|^{-1/2}
> 222\ \text{GeV}\ \text{(Muonium HFS)}.
\end{align}

\section{Dark photon signature in proposed lepton colliders\label{Sec:Colliders}}
Since the massless dark photon interactions with leptons are momentum-dependent, we should also investigate the possibility of observing LFV processes in high-energy experiments. In this section, we examine several dark-photon signatures in proposed experiments, in particular $\mu$TRISTAN~\cite{muTRISTAN2022}, a $\mu^+$-$e^-$ collider with a center of mass energy of $346~\textrm{GeV}$ and an instantaneous luminosity of $4.6\times 10^{33}\ \text{cm}^{-2}\text{s}^{-1}=
 4.6\times 10^{-6}\ \text{fb}^{-1}\text{s}^{-1}$.

Assuming the dark photon is coupled only to e-$\mu$, the two main signatures are $\mu^+ e^-\to \mu^- e^+$ (wrong sign $\mu e$ scattering), and $\mu^+ e^-\to\gamma\gamma'$ (single photon and missing energy).

\begin{figure}[t]
 \centering
 \includegraphics[width=0.6\textwidth]{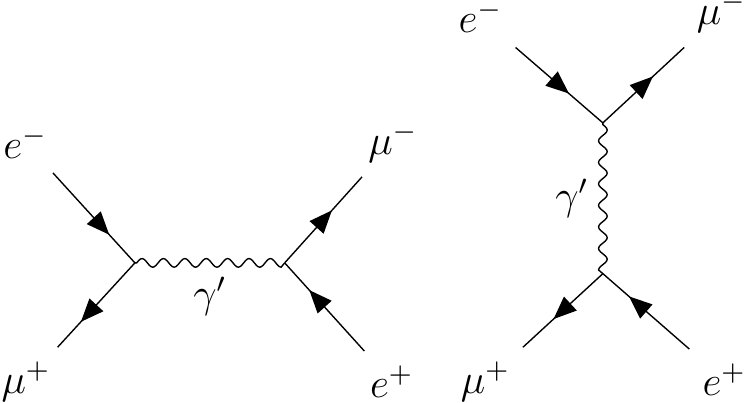}
 \caption{$s$ and $t$ diagrams involved in $\mu^+ e^-\to \mu^- e^+$.}
 \label{fig:diagrams:muetoemu}
\end{figure}

\paragraph{$\bullet\ \mu^+ e^-\to \mu^- e^+$:}
As shown in Fig.~\ref{fig:diagrams:muetoemu}, this process involves two diagrams with a tree-level dark photon exchange, in the s- and t-channel respectively. The overall cross-section reads 
\begin{align}
\sigma(\mu^+e^-\to\mu^-e^+)=
 \frac{s}{128\pi}
  \left[5
  (|D_R^{(\mu e)}|^2+|D_L^{(\mu e)}|^2)^2
        -\frac{1}{3}(|D_R^{(\mu e)}|^2-|D_L^{(\mu e)}|^2)^2\right]\,.
        \label{Eq:muetomueLimit}
\end{align}

\begin{align}\label{Eq:sigmueemu}
\sigma(\mu^+e^-\to\mu^-e^+)=
 541\left(\frac{|D_R^{(\mu e)}|}{\text{TeV}^{-1}}\right)^4\left(\frac{\sqrt{s}}{346\ \text{GeV}}\right)^2
\ \text{fb}\,,
\end{align}
in the case of  $D_L^{(\mu e)}=0$, which results in $2.5\times 10^4$ events in $10^7$ seconds for $|D_R^{(\mu e)}|=1\ \text{TeV}^{-1}$, or equivalently a scale $|{D_R^{(\mu e)}}|^{-1}\gtrsim 7~\textrm{TeV}$ requiring a minimum of 10 events in $10^7$ seconds and neglecting backgrounds (the dominant one being combinatorial or particle charge identification).

We note that the cross-section increases proportionally to $s$ owing to the dim-5 operators in Eq.~\eqref{Eq:LdipoleRL}, and eventually violates the unitarity at a larger $\sqrt{s}$. 
The amplitude is well below the unitarity bound at $\mu$TRISTAN of $\sqrt{s}=346\ \text{GeV}$.

\begin{figure}[t]
 \centering
 \includegraphics[width=0.45\textwidth]{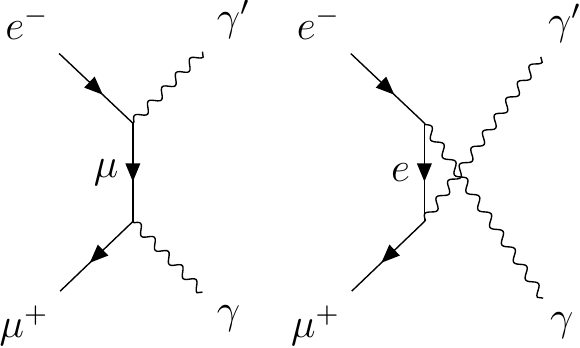}
 \caption{$t$ and $u$ diagrams involved in $\mu^+ e^-\to\gamma\gamma'$.}
 \label{fig:diagrams:muetogammagamma}
\end{figure}

\paragraph{$\bullet\ \mu^+ e^-\to\gamma\gamma'$:}
As shown in Fig.~\ref{fig:diagrams:muetogammagamma}, this process also involves two tree-level diagrams,
this time with either a muon or an electron propagator (respectively in the t- and u- channels).
The LFV coupling only appears once in these diagrams,
but the resulting enhancement is mitigated by the presence of $\alpha$ in the cross-section:
\begin{align}
\sigma(\mu^+e^-\to\gamma'\gamma)=\frac{\alpha}{2}(|D_R^{(\mu e)}|^2+|D_L^{(\mu e)}|^2)
=1521\left(\frac{\sqrt{|D_R^{(\mu e)}|^2+|D_L^{(\mu e)}|^2}}{1\ \text{TeV}^{-1}}\right)^2\ 
 \text{fb}\,.
\end{align}
This cross-section implies $7.0\times 10^4$ signal events per $10^7$ seconds at $\mu$TRISTAN for $(|D_R^{(\mu e)}|^2+|D_L^{(\mu e)}|^2)^{1/2}=1\ \text{TeV}^{-1}$.

To obtain the scale probed by this process, we have to estimate the dominant SM background: $\mu^+e^-\to\Bar{\nu}_{\mu}\nu_e\gamma$ in the case where the two neutrinos are collinear. Using Madgraph~\cite{Alwall:2014hca} with a simple model of a detector, we found that this background is proportional to the energy resolution near the $\sqrt{s}/2$ threshold. If we require a $3\sigma$ excess in the last bin assuming a $1~\textrm{GeV}$ resolution, we find that $\mu$TRISTAN could probe a scale
\begin{align}
    \frac{1}{\sqrt{|{D_{R}^{(\mu e)}}|^2+|{D_{L}^{(\mu e)}}|^2}}&\gtrsim 29 \left(\frac{\mathcal{L}}{1~\mathrm{fb}^{-1}}\right)^\frac{1}{4}~\mathrm{TeV},\\
    &\gtrsim 76~\mathrm{TeV} \text{ in $10^7$ seconds.}
\end{align}

Clearly, both constraints are orders of magnitude weaker than the one obtained by the two-body decay $\mu\to e+X$ in Eq.~\eqref{Eq:mutoegammaLimit} and Fig.~\ref{Fig:TWISTbound}.

\paragraph{$\bullet\ \mu^+e^-\to\bar\ell_1\ell_2$:}
If we allow for possible couplings of the massless dark photon to charged leptons, only three processes of the shape $\mu^+e^-\to\bar\ell_1\ell_2$ escape the constraint on the $\mu e$ coupling. They are respectively $(\bar\ell_1,\ell_2)=(\tau^+,\tau^-),(\tau^+,e^-),(\mu^+,\tau^-)$. If we neglect the  small $\mu e$ coupling, only the t-channel exchange contributes, and the cross-section for these can be expressed as
\begin{align}
\sigma&=\frac{7s}{192\pi}\left(|D_R^{(\mu\ell_1)}|^2+|D_L^{(\mu\ell_1)}|^2\right)
                        \left(|D_R^{(\ell_2 e)}|^2+|D_L^{(\ell_2 e)}|^2\right)\,,\\
      &=541\ 
        \left(\frac{\left\{\left(|D_R^{(\mu\ell_1)}|^2+|D_L^{(\mu\ell_1)}|^2\right)
                        \left(|D_R^{(\ell_2 e)}|^2+|D_L^{(\ell_2 e)}|^2\right)
                     \right\}^{1/4}}{\text{TeV}^{-1}}\right)^4
        \left(\frac{\sqrt{s}}{346\ \text{GeV}}\right)^2\ \text{fb}\,,
\end{align}
yielding the same scale as Eq.~\eqref{Eq:sigmueemu}. Once again, expected constraints by these processes can be neglected in view of other constraints:
\begin{itemize}
    \item $(\bar\ell_1,\ell_2)=(\tau^+,\tau^-)$: the coupling structures involves the product $D_{R/L}^{(\mu\tau)}D_{R/L}^{(\tau e)}$, already constrained by the limits on $\tau\to e/\mu + X$ Eq.~(\ref{Eq:tautoeormuLimit}).
    \item $(\bar\ell_1,\ell_2)=(\tau^+,e^-)$: the coupling structures involves the product $D_{R/L}^{(\mu\tau)}D_{R/L}^{(e e)}$. The first factor is already constrained by the limits on $\tau\to \mu + X$ in Eq.~(\ref{Eq:tautoeormuLimit}). The second one is also constrained as shown in Eq.~\eqref{Eq:SC}. Note that the product is also directly (weakly) constrained by $\tau\to\mu ee$, cf. Eq.~(\ref{Eq:tautomueeLimit}).
    \item $(\bar\ell_1,\ell_2)=(\mu^+,\tau^-)$: Similar to the case above with the product $D_{R/L}^{(e\tau)}D_{R/L}^{(\mu \mu)}$. cf. Eqs.~\eqref{Eq:tautoeormuLimit}, \eqref{Eq:BBN}, and \eqref{Eq:tautomumueLimit}.
\end{itemize}

Similarly, in the case of a future $e^+e^-$ collider such as the International Linear Collider (ILC)~\cite{ILCRDR2007}, the only processes that do not involve the $e\mu$ coupling while also not having huge SM backgrounds are $e^+e^-\to e\tau$ and $e^+e^-\to\mu\tau$, both already largely constrained by the stellar cooling in Eq.~\eqref{Eq:SC}, $\tau\to e/\mu + X$ in Eq.~\eqref{Eq:tautoeormuLimit}, as well as the tau three-body decays in Eqs.~\eqref{Eq:tautoeeeLimit} and \eqref{Eq:tautomueeLimit} respectively.

Overall, we find high-energy experiments to be less efficient in probing LFV couplings of the massless dark photon. Even when considering every possible family structure, All the combinations that can be probed are already constrained to a greater degree by low-energy experiments.

\section{Discussions and summary\label{Sec:Summary}}
Before summarizing our results, we briefly discuss implications on a UV model.
As mentioned in Sec.~\ref{Sec:Introduction}, a dark sector is supposed to be behind the dark photon.
One possible construction of the dark sector is the dark QED, a set of dark fermions interacting with the dark photon.
Couplings between the dark sector and the SM can be provided by a messenger sector, a set of scalars having Yukawa interaction with the dark fermions and the SM quarks and leptons.

An interesting model of this construction is proposed in Ref.~\cite{Gabrielli2016a} in order to explain the quark and lepton mass hierarchy~\cite{GabrielliRaidal2014a}.
They introduce dark fermions corresponding to the SM fermions and messenger scalars similar to squarks and sleptons.
The quark and lepton Yukawa couplings to the SM Higgs doublet are generated by one-loop diagrams with internal lines of a dark fermion and a messenger scalar.
In this way, the hierarchical chiral symmetry breaking in the dark QED with the Lee-Wick extension \cite{Gabrielli2008a} propagates to the SM sector.

The dipole interaction of the massless dark photon with the charged leptons in Eq.~\eqref{Eq:LdipoleRL} is induced at the one-loop level by the same Yukawa interaction of the messenger scalars with the SM and dark leptons. 
In principle, the messenger Yukawa interaction and thus the dipole interaction can have an arbitrary family structure. 
The one-loop contributions to the dipole couplings are characterized by the origin of chirality flip, one from the internal dark lepton mass, and another from the external SM lepton mass.
It is argued in Ref.~\cite{Gabrielli2016a} that the latter is relatively suppressed.
In this case, the dipole coupling constants are approximately written as
\begin{align}\label{Eq:UVD}
 D^{(f'f)}_R\simeq D^{(f'f)}_L\simeq c_{f'f}\frac{m_\ell}{\bar m^2_L}\,,
\end{align}
where $m_\ell$ represents the mass of the heavier charged lepton, $\bar m_L$ is the average mass of the relevant messenger scalars, and $c_{f'f}$ is a dimensionless number depending on the dark QED couplings, the messenger Yukawa couplings, the ratio of the dark lepton mass and $\bar m_L$,
and the messenger left-right mixing parameter~\cite{Gabrielli2016a}.
We note that, as stated above,  Eq.~\eqref{Eq:UVD} represents the contributions of chirality flip by the internal dark lepton mass. The charged lepton mass is radiatively generated and proportional to the dark lepton mass in the model under the present discussion. 
Hence the dark lepton mass dependence is implicitly involved in $c_{f'f}$.
Applying the model-independent result of $A=0$ in Eq.~\eqref{Eq:Jodidio} to Eq.~\eqref{Eq:UVD}, we find $\bar m_L\gtrsim \sqrt{|c_{\mu e}|}\,34\ \text{TeV}$.
As for the tau two-body decays, we obtain $\bar m_L\gtrsim \sqrt{|c_{\tau\mu(e)}|}\,5.7(5.1)\ \text{TeV}$ from Eq.~\eqref{Eq:tautoeormuLimit}.

To summarize, we have studied the leptonic interactions of the massless dark photon mainly focusing on LFV. 
Since the massless dark photon and the ordinary photon are discriminated by their interactions, the former is not coupled to the electromagnetic current.
Hence the interactions of the massless dark photon and the SM particles are described by higher-dimensional operators.
We have employed the effective dipole operators of the charged leptons as shown in Eq.~\eqref{Eq:LdipoleRL}, and examined various LFV processes in order to obtain constraints on dipole coupling constants $D_{R,L}^{(f'f)}$. 
Their inverse indicates the effective energy scale of new physics, which may be related to the mass scale of new particles if a UV model is specified as in Eq.~\eqref{Eq:UVD}.

The most stringent constraint is provided by $\mu\to e\gamma'$, $1/|D_{R,L}^{(\mu e)}|>O(10^6)\ \text{TeV}$.
Taking the angular dependence into account, we obtained the improved constraint with the result of the TWIST experiment~\cite{TWIST2015}, as presented in Table~\ref{Tab:TWIST} and Fig.~\ref{Fig:TWISTbound}.
We have also pointed out that the familon search in Ref.~\cite{Jodidio1986} gives the even stronger limit in Eq.~\eqref{Eq:Jodidio} in the case of $|D_R^{(\mu e)}|=|D_L^{(\mu e)}|$.
The angular distribution is essential to identify the chiral structure of the dipole operator and will provide valuable information on the UV theory.
If the family violating $(\mu e)$ and conserving $(ee)$ couplings coexist, $\mu\to eee$, can take place. 
The bound by the search for this process~\cite{SINDRUM1988} turns out $|D_{R,L}^{(\mu e)}D_{R,L}^{(ee)}|^{-1/2}>O(10^2)\ \text{TeV}$.
While, one finds $|D_{R,L}^{(\mu e)}D_{R,L}^{(ee)}|^{-1/2}>O(10^6)\ \text{TeV}$, combining the above constraint by $\mu\to e\gamma'$ and the astrophysical constraint of stellar cooling.

As for the LFV in tau decays, the recent search for $\tau\to\ell\gamma'$ ($\ell=\mu,e$) by Belle II~\cite{BelleII2022a} constrains the corresponding dipole couplings as $1/|D_{R,L}^{(\tau\ell)}|>O(10^3)\ \text{TeV}$.
In the presence of relevant family conserving couplings, the search for the tau decays into three charged leptons~\cite{Hayasaka2010} gives constraints, typically $|D_{R,L}^{(\tau\ell)}D_{R,L}^{(\ell'\ell')}|^{-1/2}>O(10)\ \text{TeV}$.
Again, combining the constraint by $\tau\to\ell\gamma'$ with the limit by the stellar cooling or BBN, we find $|D_{R,L}^{(\tau\ell)}D_{R,L}^{(ee)}|^{-1/2}>O(10^5)\ \text{TeV}$ and $|D_{R,L}^{(\tau\ell)}D_{R,L}^{(\mu\mu)}|^{-1/2}>O(10^4)\ \text{TeV}$.

In addition, we studied possible signatures in the recently proposed $\mu^+e^-$ collider ($\mu$TRISTAN~\cite{muTRISTAN2022}), ILC~\cite{ILCRDR2007}, and the muonium HFS measurement.
Unfortunately, we expect virtually no signals in these experiments because of the rather strong constraints by the muon decays, the tau decays, the stellar cooling, and the BBN.

In Ref.~\cite{BelleII2022a}, the Belle II experiment searched for the tau decays into a charged lepton and an invisible particle with the data of 62.8\ $\text{fb}^{-1}$.
Since their goal of integrated luminosity is about 50\ $\text{ab}^{-1}$~\cite{BelleIIPB}, we expect that Belle II will probe into the tauonic LFV of the massless dark photon at higher effective energy scales shortly.

\section*{Acknowledgment}
This work was made possible thanks to an International Research Fellowship awarded by the Japanese Society for the Promotion of Science (JSPS), under their summer program 2022.
The work of MT is supported in part by JSPS KAKENHI Grant Numbers 
JP 18K03621 and 21H00074.

\end{document}